\documentclass[aps,prd,reprint,preprintnumbers,showpacs,floatfix,nofootinbib,superscriptaddress,longbibliography]{revtex4-2}

\usepackage[utf8]{inputenc}
\usepackage{amssymb}
\usepackage{hhline}
\usepackage{amsmath}
\usepackage{mathtools}
\usepackage{siunitx}
\usepackage[dvipsnames]{xcolor}
\usepackage{array}
\usepackage{multirow,tabularx}
\usepackage{graphicx}
\usepackage{natbib}
\usepackage{xspace}
\usepackage{xstring}
\usepackage{titlesec}
\usepackage{parskip}
\usepackage[normalem]{ulem}
\allowdisplaybreaks
\usepackage{braket}
\usepackage{tikz}
\usetikzlibrary{matrix}
\usepackage{listings}
\newcommand*\diff{\mathrm{d}}
\newrobustcmd{\pea}[1]{%
	\emph{#1}\textbf{\ \ \ ---}
}
\titleformat{\paragraph}[runin]{\normalfont\normalsize\bfseries}{\emph\theparagraph}{1em}{\pea}

\newcommand*{\ie}{i.e.\@\xspace}
\newcommand*{\eg}{e.g.\@\xspace}
\newcommand*{\cf}{c.f.\@\xspace}

\newcommand*{\wrt}{w.r.t.\@\xspace}
\newcommand*{\rhs}{r.h.s.\@\xspace}

\newcommand*{\ex}{\mathrm{e}}
\newcommand*{\CLa}{$\mathcal{C}\texttt{osmo}\mathcal{L}\texttt{attice}$}

\newcommand*{\Python}{\textit{Python}\xspace}
\newcommand*{\Julia}{\textit{Julia}\xspace}

\newcommand*{\LatticeEasy}{\textit{LATTICEEASY}\xspace}
\newcommand*{\Defrost}{\textit{DEFROST}\xspace}
\newcommand*{\Hlattice}{\textit{HLATTICE}\xspace}
\newcommand*{\Gabe}{\textit{GABE}\xspace}
\newcommand*{\InflationEasy}{\textit{InflationEasy}\xspace}
\newcommand*{\GRChombo}{\textit{GRChombo}\xspace}
\newcommand*{\Stolas}{\textit{STOLAS}\xspace}

\newcommand{\solver}[1]{\texttt{#1}\xspace}
\newcommand*{\Scipy}{\textit{scipy}\xspace}
\newcommand*{\Sundials}{\textit{SUNDIALS}\xspace}
\newcommand*{\Sksundae}{\textit{sksundae}\xspace}

\usepackage{hyperref}
\hypersetup{%
     colorlinks = true,%
     linkcolor = Blue,%
     citecolor = Blue,%
     filecolor = Blue,%
     urlcolor = Blue%
     }%

\usepackage[noabbrev,nameinlink]{cleveref}
\crefname{figure}{fig.}{figs.}
\Crefname{figure}{Fig.}{Figs.}
\crefname{table}{table}{tables}
\Crefname{table}{Table}{Tables}
\crefname{section}{appendix}{appendices}
\Crefname{section}{Appendix}{Appendices}
\crefname{equation}{eq.}{eqs.}
\Crefname{equation}{Eq.}{Eqs.}

\begin{document}

\title{The limits of lattice inflation: a cautionary tale}

\author{Will Barker}
\email{barker@fzu.cz}
\affiliation{Institute of Physics of the Czech Academy of Sciences, Na Slovance 1999/2, 182 00 Prague 8, Czechia}
\affiliation{Astrophysics Group, Cavendish Laboratory, JJ Thomson Avenue, Cambridge CB3 0HE, UK}
\affiliation{Kavli Institute for Cosmology, Madingley Road, Cambridge CB3 0HA, UK}
\author{Benjamin Gladwyn}
\email{benjamin.gladwyn@physics.ox.ac.uk}
\affiliation{Department of Physics, Keble Road, University of Oxford, OX1 3RH, UK}
\author{Sebastian Zell}
\email{sebastian.zell@lmu.de}
\affiliation{Arnold Sommerfeld Center, Ludwig-Maximilians-Universit\"at, Theresienstraße 37, 80333 M\"unchen, Germany}
\affiliation{Max-Planck-Institut für Physik, Boltzmannstr. 8, 85748 Garching b.\ M\"unchen, Germany}

\begin{abstract}
Cosmological lattice simulations have become important tools for studying non-perturbative dynamics in the early Universe. Many widely used codes, however, approximate the gravitational background by an exact Friedmann--Lemaître--Robertson--Walker (FLRW) spacetime and neglect metric perturbations. We show that, during inflation, this approximation prevents the freezing of superhorizon modes. During slow roll, the curvature power spectrum decays as~$H^4$, while the deviation becomes substantially stronger during ultra-slow roll. As a result, inflationary observables can be significantly distorted. In contrast, reheating studies appear to be considerably less sensitive to the omission of metric perturbations. We propose a practical criterion for assessing the validity of FLRW simulations based on the inclusion of first-order metric perturbations, and implement it in~$\mathcal{C}\texttt{osmo}\mathcal{L}\texttt{attice}$. 
\end{abstract}

\maketitle

\paragraph*{Introduction}
In recent decades, cosmology has evolved into a precision science, driven by a remarkable increase in the accuracy of observational data. Particularly significant are high-precision measurements of the cosmic microwave background (CMB)~\cite{Planck:2018jri, BICEP:2021xfz}, which provide an unprecedented view of the early Universe more than 13 billion years ago, with upcoming next-generation experiments~\cite{Li:2017drr,SimonsObservatory:2018koc,LiteBIRD:2022cnt} expected to deliver further substantial improvements.

These measurements have led to strong observational evidence for a period of accelerated expansion in the early Universe, known as cosmic inflation~\cite{Starobinsky:1980te, Guth:1980zm, Linde:1981mu, Mukhanov:1981xt}. In the original scenario of  cold inflation, the Universe is essentially devoid of particles; once this phase has ended, a hot plasma forms through a process known as reheating.\footnote
{Alternatively, the Universe may have been warm already during inflation~\cite{Berera:1995ie}, potentially with a standard model plasma~\cite{Berghaus:2025dqi}.}
The inflationary phase stretches microscopic quantum fluctuations to macroscopic sizes which freeze upon crossing the horizon. Subsequently, the Universe enters a phase of decelerated expansion, during which the temperature decreases and the CMB is formed. The fluctuations reenter the cosmological horizon in the post-inflationary era and manifest as small perturbations in the CMB, seeding the formation of all structures in the Universe.

To fully exploit the experimental progress, theoretical predictions must achieve a comparable level of accuracy. A key conceptual limitation is that many theoretical studies rely on some form of perturbative expansion. This is particularly problematic for reheating -- an inherently non-perturbative process in which the background evolves from an inflationary phase to a radiation-dominated state. To address this challenge, lattice simulations have been employed~\cite{Khlebnikov:1996mc,Prokopec:1996rr,Khlebnikov:1996zt}, building on earlier work on topological defects~\cite{Bennett:1989yp,Allen:1990tv}.

This line of work led to the development of the public lattice code \LatticeEasy~\cite{Felder:2000hq}. Subsequently, additional tools with different ranges of applicability -- \eg in terms of admissible matter content -- were created, including \Defrost~\cite{Frolov:2008hy}, \Hlattice~\cite{Huang:2011gf}, \Gabe~\cite{Child:2013ria,Giblin:2019nuv}, and \CLa\,~\cite{Figueroa:2020rrl,Figueroa:2021yhd}. Lattice simulations have since become a standard tool for studying reheating (see~\cite{Felder:2000hj,Felder:2001kt,Garcia-Bellido:2002fsq,Easther:2006vd,Dufaux:2007pt,Lozanov:2016hid,Figueroa:2017vfa,Figueroa:2023oxc} for a selection), and there is no question they can probe regions of parameter space inaccessible to perturbative methods.

While the range of applicability of perturbation theory is well understood, the corresponding question for lattice simulations has not been addressed systematically:
\begin{center}
	\emph{What are the conditions for the validity of lattice simulations?}
\end{center}
In this letter, we initiate this line of inquiry, focusing on the following aspect. All lattice simulations mentioned above -- except for \Hlattice~\cite{Huang:2011gf} and the second version of \Gabe~\cite{Giblin:2019nuv} -- approximate the gravitational background by an exact Friedmann--Lemaître--Robertson--Walker (FLRW) metric and neglect metric perturbations. We quantify the error introduced by this approximation, both for reheating and, going further back in time, for inflation.

At first sight, lattice simulations may appear less relevant for inflation since the observed perturbations in the CMB are small, of order~$10^{-5}$. However, this expectation can be misleading. Primordial black holes~\cite{Zeldovich:1967lct,Hawking:1971ei,Carr:1974nx,Chapline:1975ojl} may form during inflation if perturbations on smaller scales -- generated later in the inflationary phase -- become sufficiently large~\cite{Starobinsky:1992ts,Dolgov:1992pu,Garcia-Bellido:1996mdl}. Moreover, small perturbations can accumulate significant effects if inflation persists for a sufficiently long time~\cite{Ford:1984hs,Antoniadis:1985pj,Starobinsky:1994bd,Tsamis:1996qq,Mukhanov:1996ak}, and even fully invalidate the description given by a classical spacetime~\cite{Dvali:2013eja,Dvali:2014gua,Dvali:2017eba}.

Indeed, inflation has been studied using lattice techniques~\cite{Caravano:2021pgc,Cai:2021hik,Caravano:2021bfn,Caravano:2022epk,Caravano:2022yyv,An:2023jxf,Caravano:2024tlp,Caravano:2024xsb,Caravano:2024moy,Sharma:2024nfu,Figueroa:2024rkr,Caravano:2025diq,Jamieson:2025ngu,Caravano:2026hca,Saha:2026cay}, and a dedicated lattice code -- \InflationEasy~\cite{Caravano:2025klk}, based on \LatticeEasy~\cite{Felder:2000hq} -- has been proposed.\footnote
{See also the code \Stolas~\cite{Mizuguchi:2024kbl} for the lattice study of stochastic inflation.}
We show that neglecting metric perturbations leads to a significant issue: it spoils the freezing of superhorizon modes which is necessary for their persistence until CMB formation. During slow roll, omitting metric perturbations causes the power spectrum decay
\begin{equation} \label{powerDecay}
	\mathcal{R}_k^2 \sim H^4 \; ,
\end{equation}
where~$H$ denotes the Hubble scale. During ultra-slow roll (USR), the deviation becomes even stronger.

In this letter, we first derive~\cref{powerDecay} and then show that the resulting decay of superhorizon modes prevents a reliable extraction of inflationary observables. We illustrate this in a simple slow-roll model with potential~$\lambda \phi^4$, which is under full perturbative control, and in a model exhibiting USR and enhanced perturbations. Furthermore, we devise a simple method to assess the relevance of metric perturbations and we demonstrate that including the \emph{first-order} metric perturbation eliminates the superhorizon decay; genuinely non-perturbative information still cannot be accessed in this setup. We implement both the diagnostic and the inclusion of first-order metric perturbations in \CLa. Importantly, we also find that reheating studies do not appear to be significantly affected by the omission of metric perturbations.

\emph{Conventions:} We work in units~$M_p\equiv 1/\sqrt{8\pi G}=1$.

\paragraph*{First-order perturbations}
We consider a scalar field~$\phi$ minimally coupled to gravity
\begin{equation} \label{action}
	S = \int{\diff^4 x\sqrt{-g} \,\Bigg[\frac{1}{2} R + \frac{1}{2}g^{\mu\nu}\partial_\mu\phi\partial_\nu\phi-V(\phi)\Bigg]}\,,
\end{equation}
and work in an FLRW background with scale factor~$a$ and conformal time~$\tau$. In the following, we use the metric
\begin{equation} \label{metric}
\begin{split}
	\diff s^2 = a^2\Big[ &-(1+ \sigma 2 A)\diff\tau^2 \\
	&+ \sigma2\partial_i B \diff x^i \diff\tau + \delta_{ij}\diff x^i \diff x^j \Big] \,,
\end{split}
\end{equation}
where~$\sigma$ is either~$0$ or~$1$. For~$\sigma=1$,~\cref{metric} includes first-order perturbations~$A$ and~$B$ in the spatially flat gauge, whereas~$\sigma=0$ is an exact FLRW background.

We split the scalar field,~$\phi = \bar{\phi} + \delta \phi$, into its background~$\bar{\phi}$ and the first-order perturbation~$\delta \phi$. As detailed in~\cref{app:perturbations}, we follow the standard procedure (see~\cite{Sasaki:1986hm, Malik:2008im, Mukhanov:2005sc, Hobson:2006se,Baumann:2022mni}) to derive the equation of motion of~$\delta \phi$, by using the Einstein equations to eliminate the metric perturbations~$A$ and~$B$, and replacing~$\mathrm{d}^2V/\mathrm{d}\phi^2$ using the background equations of motion. Defining~$f\equiv a\delta\phi$ and~$z\equiv a\bar{\phi}'/\mathcal{H}$, and transforming to momentum space, we arrive at
\begin{equation}\label{generalMS}
	f_k''+\left(k^2-\frac{z''}{z}\right)f_k= (\sigma-1) 2\varepsilon(3+\varepsilon-2\eta)\mathcal{H}^2f_k \;,
\end{equation}
where a prime denotes derivative \wrt~$\tau$ and~$\mathcal{H} \equiv a'/a$. We expressed the result in terms of the slow-roll parameters
\begin{equation}\label{slowrollparameters}
	\varepsilon\equiv1-\frac{\mathcal{H}'}{\mathcal{H}^2} \;, \qquad 	\eta \equiv 1-\frac{\bar{\phi}''}{\mathcal{H}\bar{\phi}'} \;,
\end{equation}
but it is important to note that~\cref{generalMS} does not assume any slow-roll approximation.

For~$\sigma=1$, \ie if metric perturbations are taken into account, the \rhs of~\cref{generalMS} vanishes and we obtain the well-known Mukhanov--Sasaki (MS) equation~\cite{Mukhanov:1985rz,Sasaki:1986hm}.\footnote{We show in~\cref{app:ms-numerics} that the `best' way to integrate the MS equation is \Julia's \solver{RadauIIA5} method, or \Python's \solver{Radau}.} On superhorizon scales~$k^2 \ll |z''/z|$, the solution~$f_k = z$ corresponds to a frozen curvature perturbation:
\begin{equation} \label{solutionMS}
	\mathcal{R}_k^2=\left(\mathcal{H} \frac{\delta\phi_k}{\bar{\phi}'}\right)^2=\left(\frac{f_k}{z}\right)^2 \to \text{const.}
\end{equation} 

\begin{figure*}[!t]
	\centering
	\includegraphics[width=\linewidth]{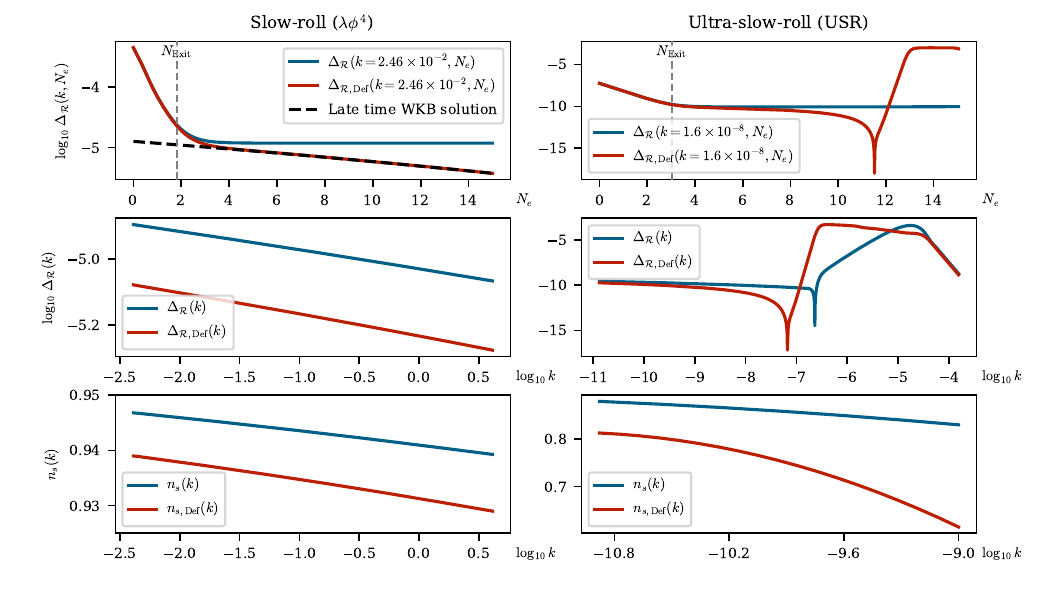}
	\caption{Deformed versus full MS equation for two models: slow-roll inflation (\cref{potential}, left) and the USR model (\cref{Jordan}, right); blue is the full result, red the deformed one throughout. \emph{Top row:} evolution of a single Fourier mode in e-folds; the frozen solution~\cref{solutionMS} is compared with the deformed MS~\cref{deformedMS} and the approximate superhorizon solution~\cref{powerDecay2} (dashed black). The slow-roll mode exits at~$N_{\mathrm{Exit}}\approx1.8$ and the USR mode at~$N_{\mathrm{Exit}}\approx3.1$, some~$10$ e-folds before the onset of USR, so it remains frozen in the MS solution but grows sharply at USR onset despite being far outside the horizon. \emph{Middle and bottom rows:} the resulting power spectrum and spectral index, each mode evaluated~$\Delta N_{\text{Exit}}=7$ e-folds after horizon crossing. In both models the deformed equation changes both significantly; for USR it enhances modes that exited long before the onset of USR.}
	\label{fig:deformedMSSolution}
	\label{fig:observablesSR}
	\label{fig:USRModeEvolution}
	\label{fig:observablesUSR}
\end{figure*}

\paragraph*{Deformed MS equation}
Now we leave out the metric perturbation, \ie take~$\sigma=0$ in~\cref{generalMS}. While it has long been known that ignoring metric perturbations changes the MS equation~\cite{Mukhanov:1988jd,Mukhanov:1990me, Hwang:1993cv}, to our knowledge the superhorizon decay of this `\emph{deformed}' MS equation under slow roll has not been studied. First, we focus on superhorizon scales~$k^2 \ll |z''/z|$ during inflation, where to leading order in slow-roll parameters~\cref{generalMS} becomes 
\begin{equation}\label{deformedMSSuperhorizon}
	f_k''-\frac{z''}{z}f_k \approx  - 6\varepsilon \mathcal{H}^2 f_k \;.
\end{equation}
In order to solve~\cref{deformedMSSuperhorizon}, we rewrite the equation in terms of e-folds~$\diff N= \mathcal{H} \diff \tau$ and define~$y \equiv \ex^\alpha f_k/z$, with~$\alpha \equiv \int\frac{1}{2}\left(3+\frac{\varepsilon_{,N}}{\varepsilon}\right)\diff N=\frac{3}{2}N+\frac{1}{2}\ln(\varepsilon)$, to obtain
\begin{equation}
	y_{,NN} - \Omega^2 y \approx 0\;,
\end{equation}
where we used~$z = a\sqrt{2\varepsilon}$ and additionally defined~$\Omega \equiv \sqrt{(\alpha_{,N})^2- 6 \varepsilon +\frac{1}{2}\left(\frac{\varepsilon_{,N}}{\varepsilon}\right)_{,N}} \approx \alpha_{,N} - 2 \varepsilon$. Consequently, the growing mode solution can be approximated as
\begin{equation}
	f_k \approx z\, \ex^{-\alpha} \ex^{\int^N \sqrt{\Omega} \, \diff \tilde{N}} \approx z\,  e^{-2\int^N \varepsilon \, \diff \tilde{N}} \;.
\end{equation}
Going back to physical time, and using~$\varepsilon = \diff\ln H/\diff N$, we arrive at~\cref{powerDecay} shown in the introduction,
\begin{equation} \label{powerDecay2}
	\mathcal{R}_k^2 \sim \left(\frac{H(N)}{H(N_{\mathrm{CMB}})}\right)^4 \;,
\end{equation}
where~$H(N_{\mathrm{CMB}})$ is the Hubble scale around horizon exit. 
In order to confirm our result, we solve the deformed MS equation,
\begin{equation}\label{deformedMS}
	f_k''+\left(k^2-\frac{z''}{z}\right)f_k= - 2\varepsilon(3+\varepsilon-2\eta)\mathcal{H}^2f_k \;,
\end{equation}
numerically without relying on any approximation. In the top-left panel of~\cref{fig:deformedMSSolution}, we show that for a slow-roll model where~$\varepsilon$ remains small, this agrees well with our analytic finding. In summary, we conclude that modes no longer freeze after horizon exit if metric perturbations are neglected. 

We can compute how~\cref{powerDecay2} changes the spectral index~$n_s \approx 1 - \diff \ln \mathcal{R}_k^2/\diff \ln N$ when we evaluate each mode at~$\Delta N_{\text{Exit}}$ e-folds after it exited the horizon:
\begin{equation}\label{DeltaNs}
	\Delta n_s = 4 \left(\epsilon(N_{\mathrm{CMB}}) - \epsilon(N_{\mathrm{CMB}} + \Delta N_{\text{Exit}})\right) \;.
	\end{equation}
This relation does not give a precise prediction of how~$n_s$ is altered, since~$N_{\mathrm{CMB}}$ is not uniquely defined. However, it is evident that the accuracy with which observables can be evaluated is inevitably limited: In order to ensure that a given mode is properly superhorizon, one must choose a sufficiently large~$\Delta N_{\text{Exit}}$, but a bigger ~$\Delta N_{\text{Exit}}$ increases the error in~\cref{DeltaNs}.

\paragraph*{Slow-roll example}
We shall first take a simple example of vanilla slow-roll inflation for a potential
\begin{equation} \label{potential}
	V(\phi) = \frac{\lambda}{4} \phi^4\;,
\end{equation}
with~$\lambda=1\times 10^{-9}$. The lower-left panels of~\cref{fig:observablesSR} show the effect of deforming the MS equation for modes initialised at~$N_\mathrm{start}$ corresponding to~$k_\mathrm{Hubble}^\mathrm{start}=k/20$, and we chose $\Delta N_{\mathrm{Exit}}=7$ as a representative value for the maximal duration typically reached in lattice simulations of inflation. We see that the power spectrum and spectral index computed using the deformed MS equation (without metric perturbations) differ significantly from the correct result due to the effect of the superhorizon decay.

\paragraph*{Ultra-slow roll example}
Next, we consider an example of a USR potential~\cite{Ballesteros:2017fsr,Ballesteros:2020qam,Barker:2024mpz}
\begin{align}
&	S = \int\diff ^4x\sqrt{-g}\biggl[\frac{1}{2}\left(1+\xi\phi^2\right)R - \frac{1}{2}g^{\mu\nu}\partial_\mu\phi\partial_\nu\phi-V(\phi)\biggr]\;, \nonumber \\
&	V(\phi) = \frac{\lambda \phi^4}{4!} \Bigg[3 + \xi^2 \phi_0^4 -8(1+c_3)\frac{\phi_0}{\phi} \nonumber \\
	&\qquad\qquad\qquad +2(1+c_2)\Big(3 + \xi\phi_0^2\Big)\frac{\phi_0^2}{\phi^2}\Bigg]\;, \label{Jordan}
\end{align}
which features an inflection point at~$\phi_0$. Following~\cite{Barker:2024mpz}, we choose~$\lambda =  6.6\times 10^{11}$,~$c_2 = 0.03$,~$c_3 = 0.075$,~$\phi_0 = 0.97$, and~$\xi = 0.10483$ to produce a significant enhancement of the power spectrum.

We first solve a single mode that exited the horizon~$10$ e-folds before the onset of the USR phase. As displayed in the top-right panel of~\cref{fig:USRModeEvolution}, the correct result corresponds to a frozen mode, but the deformed MS equation leads to a sudden drop, followed by a growth, resulting in an unphysical enhancement. The lower-right panels of~\cref{fig:observablesUSR} show that this effect also strongly alters the power spectrum and spectral index derived from it.

\begin{figure}
	\includegraphics[width=\linewidth]{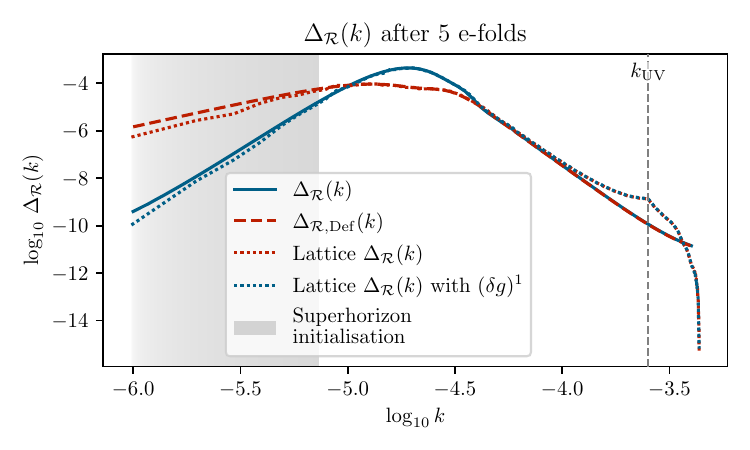}
	\caption{The power spectrum of the USR model after 5 e-folds of evolution is calculated in four ways, using the MS equation, the deformed MS equation, the lattice, and the lattice with first order metric perturbations. We observe that neglecting metric perturbations produces a large deviation in the boosting of the power spectrum. At large~$k$, we see the well-known difference of the lattice techniques from the MS and deformed MS due to the lattice UV cutoff~$k_\mathrm{UV}$ (see \cref{eq:UVcutoff}).}
	\label{fig:PSUSR}
\end{figure}

Now we come to study the power spectrum using \CLa; all details on the numerical implementation are given in~\cref{app:CL}. The evolution is initiated~$5.5$ e-folds before the end of inflation. As shown in~\cref{fig:PSUSR}, the \CLa\, result agrees well with the deformed MS equation, which differs strongly from the correct MS result. We conclude that an important reason why the lattice deviates from the MS equation is the that it does not take into account metric perturbations. Therefore, the error introduced by the FLRW approximation in lattice studies of enhanced inflationary perturbations such as~\cite{Caravano:2024tlp,Caravano:2024moy,Caravano:2025diq,Caravano:2026hca} has yet to be quantified (see details in~\cref{app:inflationEasy}).

\paragraph*{Towards a validity criterion} Importantly, deviations can occur even though the deformation term on the \rhs of \cref{deformedMS} is suppressed by the small slow-roll parameter~$\varepsilon$. 
	\begin{center}
\emph{Thus, the condition~$\varepsilon\ll1$ is not sufficient to guarantee the validity of the FLRW approximation.}
\end{center}
We demonstrate this explicitly with examples (including of USR) in~\cref{app:counterexample,app:Bayesian}, where we also show that subhorizon modes can be affected. 

It would be highly desirable to identify an \emph{a priori} criterion -- perhaps based on requiring~$\varepsilon$ to remain below a certain threshold -- that could guarantee that the error induced by neglecting metric perturbations remains small. Achieving this appears particularly challenging in the fully non-perturbative regime, where linear relations such as~\cref{generalMS,powerDecay2,DeltaNs} cease to apply and lattice simulations are most valuable. Instead, we shall now develop a method for assessing the validity of the FLRW approximation \emph{a posteriori}.

\paragraph*{Metric perturbations on the lattice}
We can compute the power spectrum of the scalar metric perturbations~$\Delta_A$ and~$\Delta_{\nabla B}$ using the quantities on the FLRW lattice, as detailed in~\cref{app:CL-AB}. We observe that metric perturbations are considerably larger for the USR example as compared to the~$\lambda\phi^4$ model (see~\cref{fig:ABk} in~\cref{app:CL-AB}). This matches our previous conclusion that the FLRW approximation causes a bigger error in former case. It is important to point out, however, that metric perturbations remain much smaller than unity even for USR. This shows that the deviation  is a cumulative effect and  small metric perturbations are a necessary but not a sufficient criterion for the validity of FLRW lattice simulations.

Different techniques of including \emph{linearized} metric perturbations on the lattice are known~\cite{Huang:2011gf,Giblin:2019nuv,Jamieson:2025ngu,Saha:2026cay}.
Through a related modification of \CLa, we can incorporate the scalar first order metric perturbation terms (see details in~\cref{app:improvedCL}).
This `improved' version of \CLa\, yields the expected freezing of superhorizon modes for the slow-roll potential (see~\cref{fig:FreezingMode,fig:PSCLvsModMSUSR} in~\cref{app:improvedCL}). Likewise, in~\cref{fig:PSCLvsModMSUSR} we show that the USR power spectrum now agrees with the behaviour of the MS equation. However, it is important to note that the lattice only becomes correct to first order in metric perturbations, just like the MS equation. Thus, the fact that all orders in field perturbations are included still cannot be exploited since the accuracy of the result is limited by the absence of higher-order metric perturbations. Only full numerical general relativity (NGR) simulations can overcome this obstacle. In summary, the different approaches include the following effects:
\begin{description}
\item[MS equation] \dotfill~$(\delta \phi)^1$ with~$(\delta g)^1$.
\item[FLRW lattice] \dotfill~$\textstyle\sum (\delta \phi)^n$ with~$(\delta g)^0$.
\item[`Improved' FLRW lattice] \dotfill~$\textstyle\sum (\delta \phi)^n$ with~$(\delta g)^1$.
\item[NGR lattice] \dotfill~$\textstyle\sum (\delta \phi)^n$ with~$\textstyle\sum (\delta g)^n$.
\end{description}

Indeed, both inflation~\cite{Clough:2016ymm,Launay:2025kef,Launay:2025lnc} and reheating~\cite{Giblin:2019nuv,Kou:2019bbc,Joana:2022uwc,Aurrekoetxea:2023jwd,Adshead:2023mvt,Adshead:2025gka,Waeming:2026wrp} have been investigated with NGR (see also review~\cite{Aurrekoetxea:2024ypv}), with some of these works employing the publicly available code \GRChombo~\cite{Andrade:2021rbd}. Evidently, the validity of lattice simulations depends on several conditions beyond the inclusion of metric perturbations, including, for example, the impact of infrared and ultraviolet cutoffs. A further issue is that lattice simulations are classical, so quantum fluctuations must be represented by classical perturbations which give a contribution to the total energy, and therefore result in a deviation in the Hubble scale (see also~\cite{Clough:2016ymm,Launay:2025kef}). We hope to revisit these questions in the future.

\paragraph*{Reheating}
FRLW lattice tools, including \CLa, have mostly been used to study reheating scenarios (see \eg~\cite{Felder:2000hj,Felder:2001kt,Garcia-Bellido:2002fsq,Easther:2006vd,Dufaux:2007pt,Lozanov:2016hid,Figueroa:2017vfa,Figueroa:2023oxc}). In \cref{app:reheating}, we show in a particular example that the inclusion of metric perturbations has no significant effect on the outcome. So in this case the FRLW approximation does not cause a relevant deviation.

While a comprehensive study of the effect of metric perturbations on reheating remains to be performed, we see at least two reasons why our example is not affected. Firstly, the relevant observables are different: total energy densities as shown in~\cref{fig:Reheating}, as opposed to properties of the spectrum of perturbations as in inflation. Secondly, we have observed relevant deviations during inflation only for superhorizon modes, while no modes exit the horizon during reheating.

\paragraph*{Conclusions} Lattice simulations are becoming an increasingly important tool in cosmology because they can access non-perturbative dynamics that lie beyond the reach of perturbative methods. Understanding their range of applicability is therefore of considerable importance. The aim of the present letter has been to initiate a systematic investigation of this question. We have focused on simulations that approximate the gravitational background by an exact Friedmann--Lema\^{i}tre--Robertson--Walker (FLRW) spacetime and neglect metric perturbations.

For inflation, we have shown that the FLRW approximation qualitatively changes the behavior of perturbations. In the absence of metric perturbations, superhorizon modes are no longer frozen. During slow-roll, this leads to the decay~\cref{powerDecay2} of the power spectrum, while in ultra-slow roll the deviation is even more pronounced. Importantly, a small slow-roll parameter~$\varepsilon\ll1$ is not sufficient to guarantee the validity of the FLRW approximation. As a result, inflationary observables extracted from FLRW simulations may differ significantly from their physical values.

First-order metric perturbations can be reconstructed from quantities available on the FLRW lattice. Their size gives an indication about the error of the FLRW approximation. Moreover, the effect of first-order metric perturbations can be incorporated into FLRW simulations, and doing so restores the expected freezing of superhorizon modes. However, if first-order metric perturbations significantly modify the result, neglecting metric perturbations is no longer justified. In that case, a treatment based solely on first-order perturbations is valid only at leading perturbative order, thereby undermining the very motivation for lattice simulations, namely the study of non-perturbative dynamics.

Our proposal is to assess the validity of FLRW simulations by recomputing the observable of interest including first-order metric perturbations, and we provide an extension of \CLa\ that allows this~\cite{supp_materials}. The absence of a significant first-order correction provides an indication that the FLRW simulation is reliable, but higher-order metric perturbations could still be important. If, however, the inclusion of first-order metric perturbations significantly modifies the outcome, numerical general relativity (NGR) is generically required to access non-perturbative information.

Our findings are intended to facilitate the use of lattice simulations by providing a practical criterion for deciding when comparatively fast FLRW simulations are sufficient and when generically more expensive NGR simulations are required. Many questions regarding the validity of lattice simulations remain open. Perhaps the most important one concerns the fact that lattice simulations are intrinsically classical, whereas the underlying theory is quantum. Quantifying the limitations imposed by the classical approximation remains an important challenge for future work.

\begin{acknowledgments}

	\paragraph*{Acknowledgments} We are grateful to Yoann Launay and Angelo Caravano for helpful discussions and thank Marco Drewes, Daniel Figueroa, Adrien Florio, Tom Giblin, Drew Jamieson, Pankaj Saha and Misha Shaposhnikov for feedback.

W.~B. is grateful for the hospitality of the Helsinki Institute of Physics and the Cavendish Laboratory at the University of Cambridge, and was supported by Marie Sk\l odowska-Curie Actions Grant Agreement No. 101081515.

The work of S.Z.~was supported by the European Research Council Gravites Horizon Grant, No. 101071779 (AO number: 850 173-6).

The work of B.G. was supported by an STFC studentship and the Clarendon Scholarship.

This research was funded in whole, or in part, by STFC, ST/Y509474/1, a Plan S funder. For the purpose of Open Access, the author has applied a CC BY public copyright licence to any Author Accepted Manuscript version arising from this submission.

This work was supported by the research environment and infrastructure of the Handley Lab at the University of Cambridge.

\textbf{Disclaimer:} Co-funded by the European Union. Views and opinions expressed are however those of the author(s) only and do not necessarily reflect those of the European Union or European Research Executive Agency. Neither the European Union nor the granting authority can be held responsible for them. 

\end{acknowledgments}

\paragraph*{Data availability} The code and data that support the findings of this article are openly available~\cite{supp_materials}.

\appendix

\begin{figure*}[t]
	\centering
	\includegraphics[width=\linewidth]{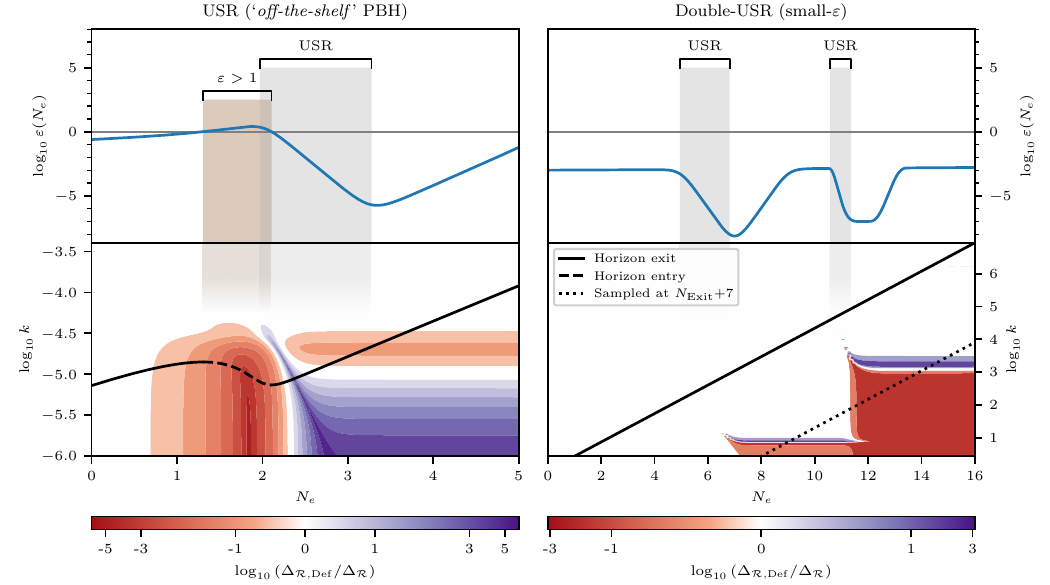}
	\caption{Breakdown of the FLRW lattice predictions for two USR models: the `\emph{off-the-shelf}' PBH model of~\cref{Jordan} (left), and the small-$\varepsilon$ example of~\cref{eq:CEmodel} (right). \emph{Top row:} the first slow-roll parameter~$\varepsilon(N_e)$; the grey line marks~$\varepsilon=1$, with the left brown band indicating the brief~$\varepsilon>1$ window, and the grey bands identifying the USR periods. \emph{Bottom row:} the ratio~$\Delta_{\mathcal{R},\mathrm{Def}}/\Delta_{\mathcal{R}}$ of the deformed (see~\cref{deformedMS}) to the true (see~\cref{generalMS}) curvature power spectrum, for each comoving mode~$k$ as a function of~$N_e$ shown in colour, with horizon exit, horizon entry, and sampling time indicated (for the `\emph{off-the-shelf}' model, all $k$ are sampled at $N_e=5$, since $N_{\text{Exit}}+7$ would extend beyond the model). Note particularly that failure even affects subhorison modes in the `\emph{off-the-shelf}' model. In the small-$\epsilon$ model shown on the right, despite~$\varepsilon<\num{1.62e-3}$ throughout, and moreover the spectrum remaining perturbative, the FLRW lattice introduces a catastrophic error before sampling is performed; the corresponding final power spectrum is shown separately in~\cref{fig:counterexample_comparison_rot}.}
	\label{fig:counterexample_comparison}
\end{figure*}

\begin{figure}[t]
	\centering
	\includegraphics[width=\linewidth]{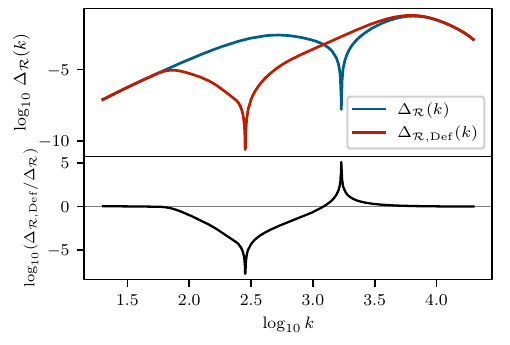}
	\caption{Final curvature power spectrum of the small-$\varepsilon$ model of~\cref{eq:CEmodel}. \emph{Top:}~$\Delta_{\mathcal{R}}(k)$ (blue) and~$\Delta_{\mathcal{R},\mathrm{Def}}(k)$ (red). \emph{Bottom:} the FLRW error~$\log_{10}(\Delta_{\mathcal{R},\mathrm{Def}}/\Delta_{\mathcal{R}})$. The FLRW lattice introduces an erroneous infrared shift of the first peak (by about a decade in $k$), and the overall spectrum is broadly wrong (with power shifted by up to five decades, over nearly two decades in $k$).}
	\label{fig:counterexample_comparison_rot}
\end{figure}

\section{Small~$\varepsilon$ guarantees nothing} \label{app:counterexample}

There are at least two conditions under which the solutions to~\cref{deformedMS,generalMS} can diverge. The first scenario is a sufficient increase in~$\varepsilon$, which modulates the magnitude of the deformation term on the \rhs of~\cref{deformedMS}. In the second scenario, the deformation term becomes dynamically significant while~$\varepsilon\ll 1$, and therefore universal bounds on the numerical value of~$\varepsilon$ have no power to prevent the failure.\footnote{A preventative bound on~$\varepsilon$ must instead be determined by first integrating the system, which fundamentally defeats the point of running simulations.} Consequently, small-$\varepsilon$ failure is more insidious, as it means that FLRW lattice simulations are not under predictive control.

We selected~\cref{Jordan} as a representative `\emph{off-the-shelf}' literature model~\cite{Barker:2024mpz}, which was in no way tuned to enhance lattice failure. The left side of~\cref{fig:counterexample_comparison} reveals a brief window where~$\varepsilon > 1$, and large-$\varepsilon$ failure arises during this window.\footnote{Worse, we notice that a significant range of \emph{sub}horizon~$k$ modes are also affected.}

We can also construct a USR model for which the FLRW lattice fails despite~$\varepsilon<\num{1.62e-3}$ throughout inflation. Rather than adopt a specific potential, we take the minimally-coupled single field of~\cref{action} and engineer its slow-roll history~$\varepsilon(N_e)$ directly (any monotonic history follows from some~$V(\phi)$ by the standard inverse construction). On a gently-rising baseline~$g_0=\num{0.03}$ (CMB pivot~$\varepsilon\simeq\num{e-3}$, so~$r=16\varepsilon\simeq0.017$) we place two localised USR episodes
\begin{equation}
\frac{\diff \ln\varepsilon}{\diff N_e} = g_0 + G_1(N_e) + G_2(N_e)\;. \label{eq:CEmodel}
\end{equation}
The first episode is a symmetric dip built from smooth top-hat functions, with rate~$r_1=7$ and edge width~$w=\num{0.4}$, defined by
\begin{equation}
\begin{aligned}
G_1 &\equiv r_1\bigl(\Sigma_{7,9}-\Sigma_{5,7}\bigr)\,, \\
\Sigma_{a,b} &\equiv \Sigma\!\left(\tfrac{N_e-a}{w}\right)\!\Sigma\!\left(\tfrac{b-N_e}{w}\right)\,, \\
\Sigma(u) &\equiv \tfrac{1}{2}\!\left(1+\tanh u\right)\,.
\end{aligned}
\label{eq:CEmodel_G1}
\end{equation}
This drives~$\varepsilon$ to~$\sim\num{e-8}$ and builds the first spectral peak. The second USR episode is an asymmetric dip. Using position~$N_2=\num{10.5}$, width~$\ell=\num{1.5}$, log-depth~$D=\num{9.6}$ and skew~$s=4$ we define this as
\begin{equation}
\begin{aligned}
G_2 &\equiv \frac{D}{\ell}\bigl(\beta_0(x_{\mathrm{r}})-\beta_s(x_{\mathrm{d}})\bigr)\,, \\
\beta_s(x) &\equiv \frac{x^2(1-x)^{2+s}}{B(3,3+s)}\,, \\
x_{\mathrm{d}} &\equiv \frac{N_e-N_2}{\ell}\,, \\
x_{\mathrm{r}} &\equiv x_{\mathrm{d}} - 1\,.
\end{aligned}
\label{eq:CEmodel_G2}
\end{equation}
Here~$B$ is the Euler beta function and the bumps~$\beta_s(x)$ vanish outside~$x\in[0,1]$. It is the skew~$s>0$ that lets this second USR episode deform the first peak, while its own peak stays perturbative.

The model in~\cref{eq:CEmodel} is shown on the right side of~\cref{fig:counterexample_comparison}, where the introduction of two USR episodes on top of an underlying slow-roll baseline is visible. The final power spectrum is given in~\cref{fig:counterexample_comparison_rot}, which shows how the FLRW lattice is unable to correctly characterise the power enhancements corresponding to these USR episodes. The small-$\varepsilon$ model in~\cref{eq:CEmodel} makes it especially clear that the~$\varepsilon\ll 1$ condition cannot be used as a reliability criterion for science products in lattice simulations of inflation.

\section{Universality of lattice failure} \label{app:Bayesian}

\begin{figure*}[t]
	\centering
	\includegraphics[width=\linewidth]{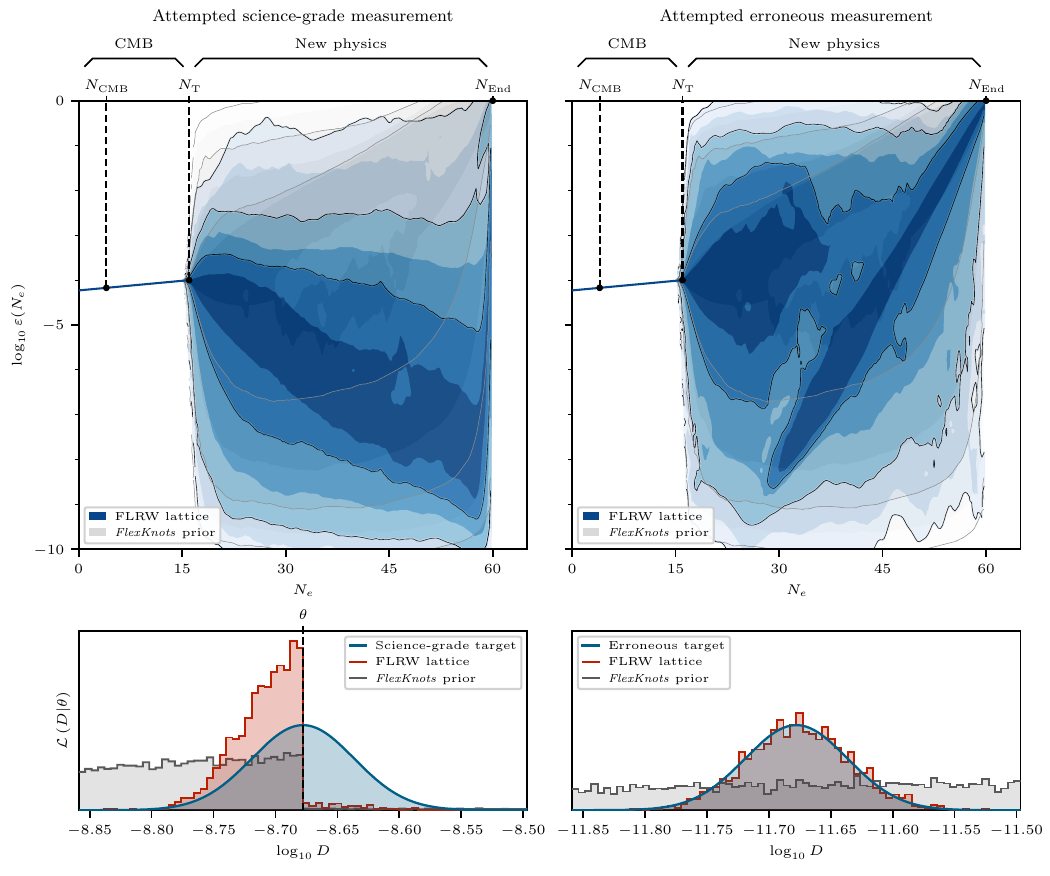}
	\caption{Strong constraints on the inflationary model are necessary for the FLRW lattice to measure the CMB power spectrum. \emph{Top:} non-parametric functional posteriors (blue) and priors (gray) of the first slow-roll parameter~$\varepsilon(N_e)$, including the known CMB epoch and possible new physics that follows, as accommodated by~\cref{eq:genmodel,eq:fconds}. \emph{Bottom:} likelihood of the FLRW lattice measuring the data~$D\equiv\Delta_{\mathcal{R},\mathrm{Def}}(k_{\mathrm{CMB}})$, when attempting to find the true parameter value~$\theta\equiv \Delta_{\mathcal{R}}(k_{\mathrm{CMB}})\simeq\num{2.1e-9}$ at the end of inflation, for a fixed mid-CMB mode~$k_{\mathrm{CMB}}=a(N_{\mathrm{CMB}})H(N_{\mathrm{CMB}})$ with~$N_{\mathrm{CMB}}=4$. Shown are the target likelihoods in~\cref{eq:targetlike,eq:wronglike} (blue), and what the FLRW lattice achieves (red), along with the functional priors (gray). \emph{Left:} when attempting to perform a science-grade measurement of the CMB power (a broad Gaussian in~\cref{eq:targetlike}, centred on the true value), the FLRW lattice requires that all new physics be strongly constrained. Moreover, the result is still biased by power suppression. \emph{Right:} when CMB power suppression is actively sought (the same Gaussian in~\cref{eq:wronglike}, now centred on one thousandth of the correct value), the FLRW lattice performs well. Consequently, FLRW lattice simulations are capable of predicting specific and wrong CMB physics, so long as the inflationary phenomenology that is left unconstrained by the CMB is first sacrificed.}
	\label{fig:flex}
\end{figure*}

If FLRW lattice simulations are to be useful as probes of inflation, they should at least demonstrate the ability to measure the known CMB band of the power spectrum when inflation ends. Such a measurement presumes an FLRW lattice evolving over $50$--$60$ e-folds, which is beyond current capabilities. Yet the alternative technique --- sampling at seven e-folds after horizon exit --- itself presumes that superhorizon modes are correctly frozen, and by now it is clear that this is not the case. We can therefore use~\cref{generalMS,deformedMS} to explore what a measurement at the end of inflation would look like, at the formal level without a real lattice. A general assessment is beyond the scope of tuned examples, such as those studied in~\cref{app:counterexample}: rather, it requires a Bayesian approach. In this appendix it is shown that the FLRW lattice is generally \emph{unable} to measure the CMB power spectrum, and that attempts at such a measurement must constrain any inflationary phenomenology that lies above CMB scales.

The first step is to establish a very general instance of single-field inflation, consistent with CMB observations~\cite{Planck:2018jri,BICEP:2021xfz}. This may be written as
\begin{equation}\label{eq:genmodel}
	\varepsilon(N_e)\equiv
	\begin{cases}
		\varepsilon_0\mathrm{e}^{(1-n_s)(N_e-N_{\mathrm{T}})}\,, & 0\leq N_e\leq N_{\mathrm{T}}\,,\\[2pt]
		f(N_e)\,, & N_{\mathrm{T}}\leq N_e\leq N_{\mathrm{End}}\,.\\[2pt]
	\end{cases}
\end{equation}
In~\cref{eq:genmodel}, there are two distinct epochs. During the initial CMB epoch,~$\mathrm{d}\ln\varepsilon/\mathrm{d}N_e=1-n_s-2\varepsilon\simeq1-n_s$ supplies the observed red tilt. Beyond the transition point~$N_{\mathrm{T}}$ there follows an epoch of new physics which is not constrained by the CMB, and which is functionally described by
\begin{equation}\label{eq:fconds}
\begin{gathered}
	f(N_{\mathrm{T}})\equiv\varepsilon_0,\qquad f(N_\mathrm{End})\equiv 1,\\
	\varepsilon_{\mathrm{Min}}\leq f(N_e)\leq 1\,.
\end{gathered}
\end{equation}
Thus,~\cref{eq:fconds} provides continuity with the CMB and (at least) one opportunity for a graceful exit at~$N_\mathrm{End}$.

One reasonable choice of parameters takes~$n_s=0.967$, and (notwithstanding the results of~\cref{app:counterexample}) it is simplest to select~$\varepsilon_0=\varepsilon(N_{\mathrm{T}})=\num{e-4}$ so that the CMB power spectrum is at least produced during the small-$\varepsilon$ regime. To ensure a compact prior in what follows, all new physics is bounded to lie above~$\varepsilon_{\mathrm{Min}}=\num{e-10}$ between its onset at~$N_{\mathrm{T}}=16$ and the nominal end of inflation at~$N_{\mathrm{End}}=60$.

In a very simple scenario, we care only about the true value of some fiducial point in the CMB power spectrum. This can be understood as a `\emph{model}' with one parameter~$\theta\equiv \Delta_{\mathcal{R}}(k_{\mathrm{CMB}})$ measured at~$N_\mathrm{End}$. For a fixed mid-CMB mode~$k_{\mathrm{CMB}}=a(N_{\mathrm{CMB}})H(N_{\mathrm{CMB}})$ exiting at~$N_{\mathrm{CMB}}=4$, this is
\begin{equation}\label{eq:theta}
	\theta=\Delta_{\mathcal{R}}(k_{\mathrm{CMB}})\simeq\num{2.1e-9}\,.
\end{equation}
The FLRW lattice simulation acts as an `\emph{experiment}' which measures the value of~$\theta$ as the single `\emph{data}' point~$D\equiv\Delta_{\mathcal{R},\mathrm{Def}}(k_{\mathrm{CMB}})$. A perfect experiment would have an ideal `\emph{likelihood}' function~$\mathcal{L}(D|\theta)=\delta(D-\theta)$, but for the FLRW lattice it is known that there is a systematic error due to the evolution of~$\varepsilon(N_e)$ between~$N_{\mathrm{CMB}}$ and~$N_\mathrm{End}$. Actually, due to the choice of parameters, it can be shown that almost all of this error now depends on~$f(N_e)$, and accumulates during the new physics epoch.\footnote{When checking this,~\cref{app:counterexample} makes it clear that the small-$\varepsilon$ regime of the CMB epoch is a meaningless criterion. It must be confirmed directly that~$\log_{10}\left[\Delta_{\mathcal{R},\mathrm{Def}}(k_{\mathrm{CMB}})/\Delta_{\mathcal{R}}(k_{\mathrm{CMB}})\right]\simeq\num{-1.8e-3}$ when measured at~$N_\mathrm{T}$. In terms of the (more generous) requirement to be imposed in~\cref{eq:targetlike}, this error lies below~$\num{0.05}\sigma$.}

Despite this error, one can ask if it is still possible to produce an efficacious (i.e. science-grade)~$\mathcal{L}(D|\theta)$ by running FLRW lattice simulations across an appropriate ensemble of~$f(N_e)$ histories. A concrete example is
\begin{equation}\label{eq:targetlike}
	\mathcal{L}(D|\theta)=\frac{1}{\sqrt{2\pi}\sigma}\exp\left[-\frac{1}{2\sigma^2}\left(\log_{10}\frac{D}{\theta}\right)^2\right],
\end{equation}
where~$\sigma=\log_{10}\left(1.1\right)\simeq\num{0.041}$, so that a one-sigma discrepancy corresponds to the FLRW lattice over- or under-predicting the CMB power by a factor of~$1.1$.\footnote{This is not an unreasonable target: the power in the lowest-$k$ shells of a lattice is averaged over only~$M\sim6$ modes and so fluctuates at the~$1/\sqrt{M}\sim40\%$ level, so~\cref{eq:targetlike} presumes an ensemble of $\mathcal{O}(30)$ realisations.} Such a likelihood function would typically be used to infer~$\theta$ for a fixed~$D$. Equally, and for fixed~$\theta$,~\cref{eq:targetlike} also implies some distribution of~$D$ values, and hence a distribution of~$f(N_e)$ histories over which FLRW lattice simulations must be marginalised if the desired likelihood is to be produced.

The~$f(N_e)$ distribution admits a non-parametric reconstruction by means of \emph{FlexKnots}~\cite{Handley:2019fll,Ormondroyd:2025exu,Ormondroyd:2025iaf,Ormondroyd:2025phk,Kroupa:2026eqg,Kroupa:2026xda}. In this scheme,~$f(N_e)$ is modelled as a piecewise-linear interpolation through~$n$ interior knots, drawn uniformly from the domain and range implied by~\cref{eq:genmodel,eq:fconds}. The resulting~$\varepsilon(N_e)$ is then smoothed via the mollification
\begin{equation}\label{eq:smoothing}
	\frac{\mathrm{d}\log_{10}\varepsilon}{\mathrm{d}N_e}\to\int\mathrm{d}N_e'\,K\left(N_e,N_e'\right)\frac{\mathrm{d}\log_{10}\varepsilon\left(N_e'\right)}{\mathrm{d}N_e'}\,,
\end{equation}
where~$K\left(N_e,N_e'\right)\propto\exp\left[-\left(1-u^2\right)^{-1}\right]$ for~$|u|<1$ and vanishes otherwise, with~$u=\left(N_e-N_e'\right)/w\left(N_e'\right)$, and where the width~$w(N_e)=\min\left(\Delta,N_e,N_\mathrm{End}-N_e\right)$ tapers to zero at either end of the domain, with~$\Delta=1.13$, so that~\cref{eq:smoothing} renders~$\varepsilon(N_e)$ continuously differentiable and $\eta'$ contains no delta functions. Moreover,~$K$ is normalised in its first argument, while its width is set by the second, so~\cref{eq:smoothing} conserves~$\int\mathrm{d}N_e\,\mathrm{d}\log_{10}\varepsilon/\mathrm{d}N_e$ exactly and the terminal condition~$f(N_\mathrm{End})\equiv 1$ of~\cref{eq:fconds} survives the smoothing.\footnote{Note that~\cref{eq:smoothing} encroaches slightly on the CMB epoch. We checked that this has negligible effect on the results which follow.} The number and positions of knots is nested-sampled~\cite{Skilling:2004pqw,Skilling:2006gxv,Handley:2015fda}.\footnote{We use $n\leq 8$ interior knots, with positions drawn from a sorted uniform prior on $N_e$ and heights drawn from a uniform prior on $\log_{10}\varepsilon$. Sampling is performed with \emph{PolyChord}~\cite{Handley:2015fda}, with~$\texttt{num\_live}=1000$ live points and~$\texttt{num\_repeats}=34$ (twice the dimensionality of the parameter space), and clustering disabled. All the priors shown in~\cref{fig:flex} are computed independently by Monte Carlo sampling. See supplemental materials at~\cite{supp_materials} for full numerical details.}

After sampling, the left column of~\cref{fig:flex} shows that the functional posterior of~$f(N_e)$ has been tightly constrained. Despite this apparent activity, the reconstructed~$\mathcal{L}(D|\theta)$ remains biased (power-suppressed) relative to the target in~\cref{eq:targetlike}. The bias is prior-driven: it is evidently hard to draw a~$f(N_e)$ trajectory from the current prior that \emph{enhances} the power at~$N_\mathrm{End}$ without overshooting. Two conclusions follow:
\begin{enumerate}
	\item To attempt a correct measurement of the CMB power spectrum using an FLRW lattice, it is first necessary to strongly constrain the inflationary model. This limits the capacity of the FLRW lattice to explore new physics.
	\item Moreover, it is hard to find a constrained choice of inflationary model that gives a correct measurement, with most choices leading to suppression of CMB power.
\end{enumerate}

With the same infrastructure, it is also possible to ask for an FLRW lattice likelihood that is catastrophically wrong, by switching the target from~\cref{eq:targetlike} to
\begin{equation}\label{eq:wronglike}
	\mathcal{L}(D|\theta)=\frac{1}{\sqrt{2\pi}\sigma}\exp\left[-\frac{1}{2\sigma^2}\left(\log_{10}\frac{D}{\theta}-s_0\right)^2\right],
\end{equation}
with~$s_0=-3$ and~$\sigma$ unchanged. The target in~\cref{eq:wronglike} now asks the FLRW lattice to precisely under-predict the CMB power by a factor of~$10^3$. The result is shown on the right of~\cref{fig:flex}. This time, the target is well achieved by the FLRW lattice.

Certain features of~\cref{fig:flex} also support the findings of~\cref{app:counterexample}. The left plot, where accuracy is sought, resists exiting inflation until the last possible moment. This may be understood as an avoidance of the large-$\varepsilon$ error scenario. By contrast the right plot is less averse to large~$\varepsilon$, but nor does it stick to~$\varepsilon\simeq 1$ in order to produce the sought error. Instead, a sudden cliff at around $N_e\simeq 28$ plays a role in the failure, while the left plot prefers to explore small-$\varepsilon$ regions cautiously, via a gentle slope. This appears consistent with fact that the~$\varepsilon(N_e)$ model investigated in~\cref{app:counterexample} achieves small-$\varepsilon$ failure through USR. These observations are, however, qualitative, and establishing their generality is beyond the scope of the present work.

The CMB power spectrum is a natural focus because it is well constrained by observations. This choice is illustrative, however, and the problem it reveals is more pervasive. The analysis can straightforwardly be extended to measure the power at smaller scales, by postponing~$N_{\mathrm{T}}<N_\mathrm{End}$ to some later time, and backfilling with specific new physics scenarios. There is no reason to think that this will lead to a more favourable outcome.

The choice of \emph{FlexKnots} allows for many improvements. For example, one could place the piecewise-linear spline on the second derivative of~$\log_{10}\varepsilon$ and then integrate twice. This removes any need for the convolution in~\cref{eq:smoothing}, and a similar approach was used in~\cite{Handley:2019fll} to reconstruct the inflationary potential from \emph{Planck} 2018 data.\footnote{Such alternatives affect the computational cost, and may reconfigure the functional prior on~$f(N_e)$. It would be interesting to understand whether this alleviates the bias on the lower left plot of~\cref{fig:flex}.} Indeed, a more principled approach fully relaxes~\cref{eq:genmodel,eq:fconds}, and reconstructs the CMB epoch jointly by using the actual observations.

\section{Single-field first-order perturbations} \label{app:perturbations}
Varying the action in~\cref{action} results in the Klein--Gordon equation,
\begin{align}
	\begin{split}
	\delta\phi''+2\mathcal{H}\delta\phi'-\nabla^2\delta\phi=&\sigma(A'+\nabla^2B)\bar{\phi}'\\&-2a^2V_{,\phi}\sigma A-a^2V_{,\phi\phi}\delta\phi\,,
	\end{split}
\end{align}
where~$V_{,\phi}\equiv \mathrm{d}V/\mathrm{d}\phi$ and~$V_{,,\phi}\equiv \mathrm{d}^2V/\mathrm{d}\phi^2$. Using the Einstein equations, we can eliminate the metric perturbations in favour of~$\delta\phi$ with
\begin{subequations}
\begin{align}
	A &= \varepsilon \frac{\mathcal{H}}{\bar{\phi}'}\delta\phi\,, \label{eq:A} \\
	\nabla^2 B &= -\varepsilon \frac{\mathcal{H}}{\bar{\phi}'} \left(\delta\phi'+(\eta-\varepsilon)\mathcal{H}\delta\phi\right)\,, \label{eq:B}
\end{align}
\end{subequations}
for slow-roll parameters defined in~\cref{slowrollparameters}, but where no slow-roll approximation has been made. We obtain
\begin{equation}\label{eq:metricPerturbation}
\begin{aligned}
	& \delta\phi''+2\mathcal{H}\delta\phi'-\nabla^2\delta\phi = \\
	&\qquad \; \big[\underbrace{2\sigma\varepsilon(3+\varepsilon-2\eta)}_\text{Metric PT terms}-\frac{a^2V_{,\phi\phi}}{\mathcal{H}^2}\big]\mathcal{H}^2\delta\phi\,,
\end{aligned}
\end{equation}
where the underbrace indicates the terms that originated from the metric perturbation. We eliminate~$V_{,\phi\phi}$ using the background equation of motion
\begin{equation}
	- \frac{a^2 V_{,\phi\phi}}{\mathcal{H}^2} = (\eta - 3)(\varepsilon + \eta) - \frac{\eta'}{\mathcal{H}}\,,
\end{equation}
and define~$f\equiv a\delta\phi$ to eliminate the friction term~$2\mathcal{H}\delta\phi'$ to give
\begin{equation}
	\delta\phi''+2\mathcal{H}\delta\phi' = \frac{1}{a}\left(f''-(2-\varepsilon)\mathcal{H}^2 f \right)\,.
\end{equation}
Therefore, this becomes
\begin{align}
	\begin{split}
	& f''-(2-\varepsilon)\mathcal{H}^2 f -\nabla^2 f = \big[2\sigma\varepsilon(3+\varepsilon-2\eta)\\&\qquad\qquad +(\eta - 3)(\varepsilon + \eta) - \frac{\eta'}{\mathcal{H}}\big]\mathcal{H}^2 f\,.
	\end{split}
\end{align}
Defining~$z\equiv \frac{a\bar{\phi}'}{\mathcal{H}}$, we have
\begin{equation}
	\frac{z''}{z} = \left[ 2 + 2\varepsilon - 3\eta + (2\varepsilon - \eta)(\varepsilon - \eta) - \frac{\eta'}{\mathcal{H}} \right] \mathcal{H}^2\,,
\end{equation}
and so for the Fourier mode~$f_k = \int f(\tau, \bf{x})e^{-i\bf{k}\cdot\bf{x}}\diff^3x$, we arrive at~\cref{generalMS}.

\section{Lattice implementation} \label{app:CL}

For the USR model we choose a lattice size of~$N^3=512^3$, a program time step of~$\texttt{dt}=100$, and~$f_*=\omega_*=2.435\times10^{18}$. The IR cutoff~$k_\mathrm{IR} = 9.8\times 10^{-7}$ defines the side length of the comoving box via~$k_\mathrm{IR}=2\pi/L$. The UV cutoff is given by the grid discretization,
\begin{equation}\label{eq:UVcutoff}
	k_\mathrm{UV}\equiv \frac{N}{2}k_\mathrm{IR}= \frac{\pi}{\delta x},
\end{equation}
for lattice spacing~$\delta x = L/N$, whilst the maximum momentum is~$k_\mathrm{max}=\sqrt{3}k_\mathrm{UV}$. To mimic quantum fluctuations, \CLa\, imposes vacuum fluctuations on the scalar fields as Gaussian random fields in momentum space, for more see~\cite{Figueroa:2020rrl, Figueroa:2021yhd}. 

For the slow-roll model, we use~$N^3=64^3$,~$\texttt{dt}=0.00001$,~$\omega_*=1.64\times10^{15}$,~$f_*=5.20\times10^{19}$, and~$k_\mathrm{IR} = 1.54\times 10^{-3}$. For the reheating model, we use~$N^3=64^3$,~$\delta\eta=0.003/\omega_*$,~$\omega_*=1.56\times10^{13}$,~$f_*=2.04\times10^{18}$, and~$k_\mathrm{IR} = 1.92\times 10^{-5}$. For all three models we use the~$\texttt{VV2}$ solver. See supplemental materials at~\cite{supp_materials} for full numerical details.

\section{Comment on \InflationEasy} \label{app:inflationEasy}

\begin{figure}
	\centering
	\includegraphics[width=\linewidth]{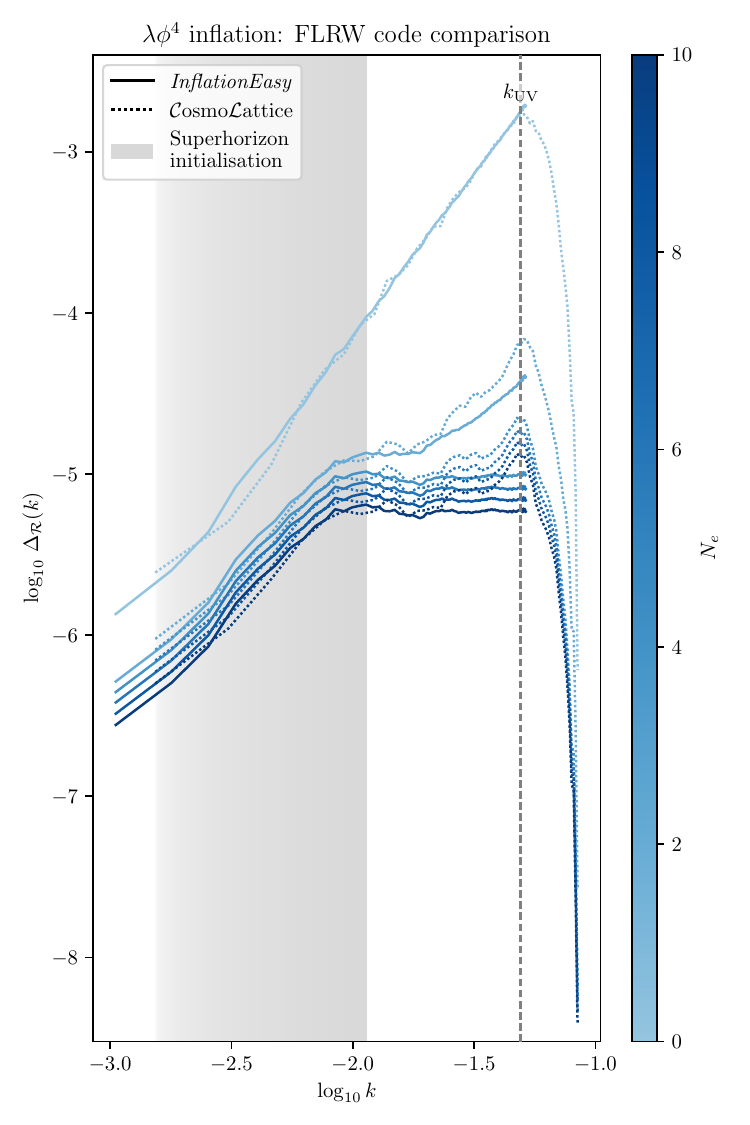}
	\caption{The curvature power spectrum~$\Delta_\mathcal{R}(k)$ of the~$\lambda\phi^4$ slow-roll model computed with \InflationEasy (solid) and \CLa{} (dotted), with the colour encoding the number of e-folds~$N_e$. The two FLRW codes agree across the resolved range, confirming that they share the same approximation. In particular, the superhorizon power spectrum is not frozen. The sharp fall-off of the \CLa\, curves towards~$k_\mathrm{UV}$ is the lattice ultraviolet cutoff, which \InflationEasy handles more gracefully.}\label{fig:inflationEasy}
\end{figure}

The lattice studies~\cite{Caravano:2024tlp,Caravano:2024moy,Caravano:2025diq,Caravano:2026hca} of scalar single-field inflation (and~\cite{Caravano:2024xsb} of axion inflation) have been performed with the code \InflationEasy~\cite{Caravano:2025klk} (or predecessor versions). \InflationEasy relies on the same FLRW approximation as \CLa, and accordingly the results of the two codes agree, as shown in~\cref{fig:inflationEasy}.

The inflationary models in~\cite{Caravano:2024tlp,Caravano:2024moy,Caravano:2025diq,Caravano:2026hca} are chosen to have a significantly smaller~$\varepsilon$ than the two examples in the main part of the present paper.\footnote
{The theoretical consistency of the~$\eta$-parametrization used in \cite{Caravano:2024moy,Caravano:2025diq,Caravano:2026hca} has recently been called into question~\cite{Barnert:2026ijs}. We also note that in~\cite{Caravano:2024tlp,Caravano:2024xsb,Caravano:2024moy,Caravano:2025diq,Jamieson:2025ngu,Caravano:2026hca} numerical results are mostly compared to analytic findings derived in the absence of metric perturbations (\cf~\cite{Inomata:2022yte,Franciolini:2023agm,Ballesteros:2024zdp}).}
We have shown in~\cref{app:counterexample}, however, that catastrophic error can arise perturbatively when the metric perturbations are arbitrarily excluded, due to dyanmical effects that are strictly independent of the smallness of~$\varepsilon$. The situation is even worse in the fully non-perturbative regime when the linear~\cref{generalMS} is no longer valid and so developing a criterion for the validity of the FLRW-approximation becomes even harder.

\section{Spectrum of lattice metric perturbations} \label{app:CL-AB}

\begin{figure}
	\includegraphics[width=\linewidth]{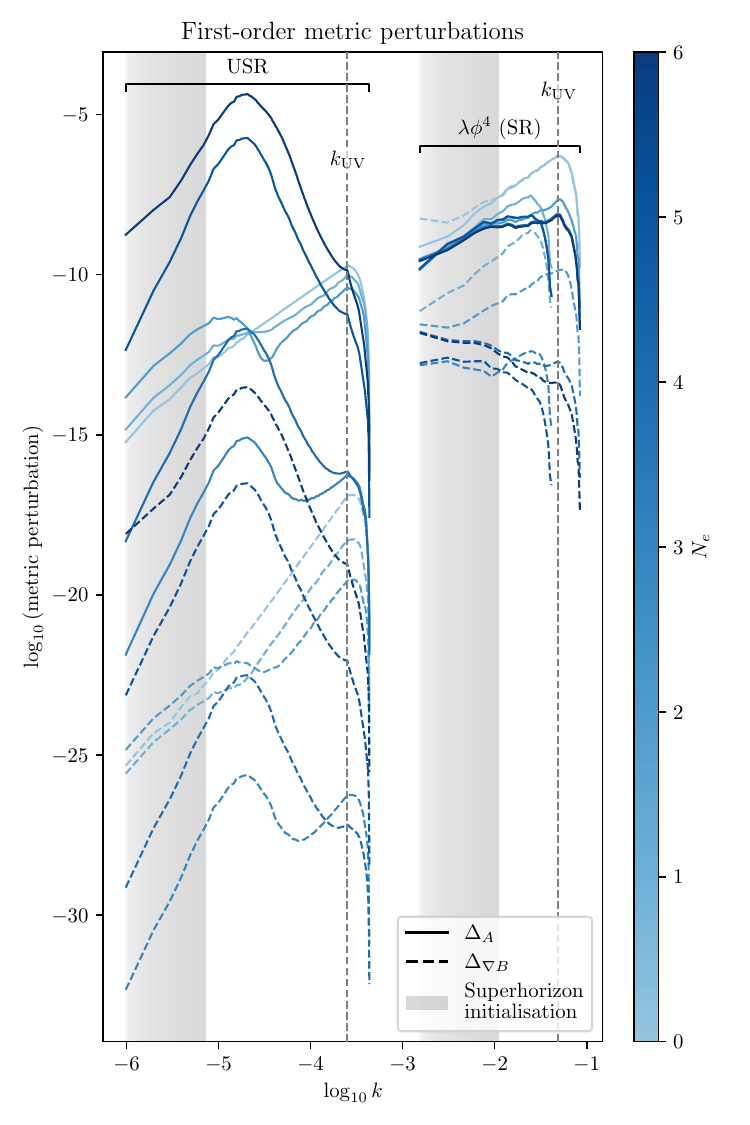}
	\caption{The magnitude of the first order metric perturbations for the two inflationary models. Metric perturbations are significantly larger for the USR example as compared to the slow roll model. Nevertheless, superhorizon secular effects of the metric perturbation persist in both cases and invalidate the result.}\label{fig:ABk}
\end{figure}

The power spectrum of metric perturbations can be estimated from the power spectra of the field fluctuations on the lattice. Recall that the dimensionless power spectrum~$\Delta_\phi(k)$ is defined by
\begin{equation}
    \mathcal{P}_\phi(k) \equiv \frac{2\pi^2}{k^3} \Delta_\phi(k)\,,
\end{equation}
\begin{equation}
    \langle \delta \phi_{\mathbf{k}} \delta \phi_{\mathbf{k}'} \rangle = (2\pi)^3 \mathcal{P}_{\phi}(k)\delta^{(3)}(\mathbf{k}+\mathbf{k}')\,.
\end{equation}
The lapse perturbation at wavenumber~$k$ is
\begin{equation}
    A_k = \varepsilon \frac{\mathcal{H}}{\bar{\phi}'}\delta\phi_k\,,
\end{equation}
and so the power spectrum of the lapse perturbation is
\begin{equation}
    \Delta_A = \left( \varepsilon \frac{\mathcal{H}}{\bar{\phi}'}\right)^2 \Delta_\phi\,.
\end{equation}
For the shift perturbation at~$k$ we have 
\begin{equation}
	k^2B_k = -\varepsilon \frac{\mathcal{H}}{\bar{\phi}'} \left(\delta\phi'_k+(\eta-\varepsilon)\mathcal{H}\delta\phi_k\right)\,,
\end{equation}
and an upper bound on the power spectrum is
\begin{equation}
    \Delta_{\nabla B} = \frac{1}{k^2}\left( \varepsilon \frac{\mathcal{H}}{\bar{\phi}'}\right)^2 \left(\Delta_{\phi'}+(\eta-\varepsilon)^2\mathcal{H}^2\Delta_{\phi}\right)\,.
\end{equation}
The FLRW approximation requires~$\Delta_A \ll 1$ and~$\Delta_{\nabla B} \ll 1$ for all relevant modes; when either condition is violated for a given mode, the metric perturbation for that mode is no longer small and the FLRW lattice is immediately inconsistent. Nevertheless, even when~$\Delta_A$ and~$\Delta_{\nabla B}$ are small, the affect of metric perturbations may not be small, as observed for the slow-roll and USR examples displayed in~\cref{fig:ABk}.

\section{First-order metric perturbations on the lattice} \label{app:improvedCL}

\begin{figure}
	\includegraphics[width=\linewidth]{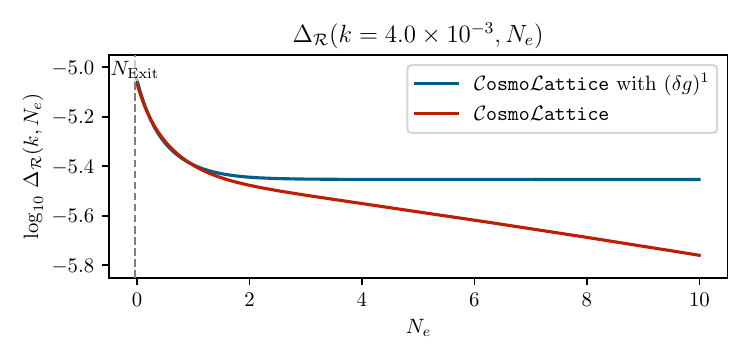}
	\caption{The `improved' version of \CLa\, includes the first order metric perturbations. For the slow-roll model, we recover -- unlike in the top-left panel of~\cref{fig:deformedMSSolution} -- the freezing behaviour of the superhorizon modes expected from the MS equation. The plotted mode crosses the horizon at~$N_e=N_{\mathrm{Exit}}\approx0$, so it is superhorizon throughout the evolution shown.}\label{fig:FreezingMode}
\end{figure}

\begin{figure}
	\centering
	\includegraphics[width=\linewidth]{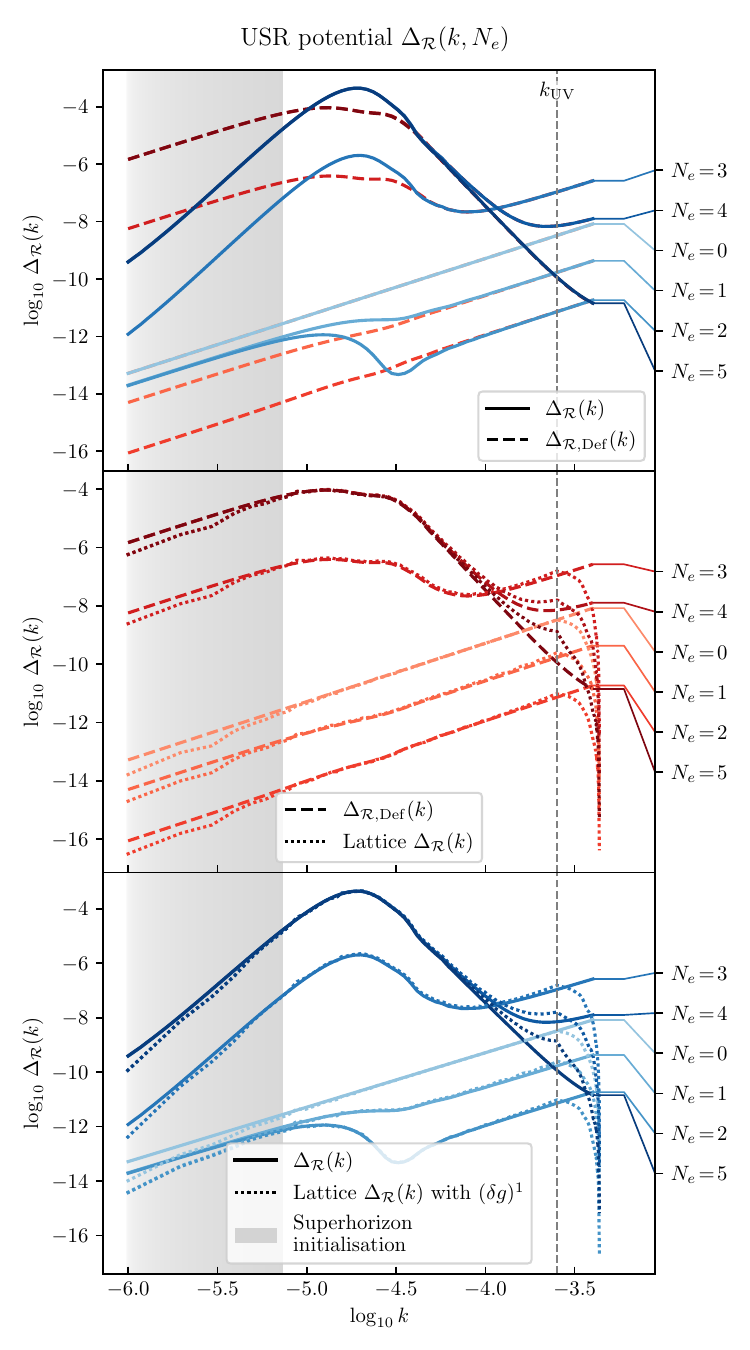}
	\caption{\emph{Top:} The full and deformed MS power spectra for the USR model. Neglecting metric perturbations produces a large difference in the boosting of the power spectrum. \emph{Middle:} The evolution of the power spectrum in \CLa\, (dotted) compared to the power spectrum expected from the deformed MS equation (dashed) -- the lattice tracks the deformed analytic, confirming that the dominant error from omitting metric perturbations on the lattice is captured by the deformed MS equation. \emph{Bottom:} The `improved' version of \CLa\, includes the first-order metric perturbations (dotted); the power spectrum now matches the full MS result (solid) closely. In the two bottom plots, deviations on the left are caused by the fact that modes are not initialized (sufficiently) inside the horizon, while on the right we see the effect of the lattice UV-cutoff~$k_\mathrm{UV}$.}
	\label{fig:PSCLvsModMSUSR}
\end{figure}

The field dynamics in an expanding background are solved on a lattice using a discrete version of the continuum equation of motion and iterating for some finite number of time steps. Varying the action gives
\begin{equation}
	(a^3\phi')'-a\nabla^2\phi+a^3V_{,\phi}=0
\end{equation}
so that when expanding the first term, the standard second derivative and friction terms become explicit. Instead of expanding to a second derivative, we may write as two equations at first order. Labelling the conjugate momenta of~$\phi$ as~$\pi\equiv\phi'$, we have
\begin{subequations}
\begin{align}
	\phi^{\prime} &= a^{-3} \pi, \\
	\pi^{\prime} &= a \nabla^2 \phi - a^3 V_{, \phi}.
\end{align}
\end{subequations}
The same procedure for the scale factor~$\pi_a\equiv a'(\tau)$ gives us the system of equations
\begin{subequations}
\begin{align}
	\pi_a(\tau) &= a^{\prime}(\tau), \\
	\pi_a^{\prime}(\tau) &= \mathcal{K}_a[a(\tau), E_V(\tau), E_K(\tau), E_G(\tau)], \\
	\pi(\mathbf{x}, \tau) &= \mathcal{D}[\phi^{\prime}(\mathbf{x}, \tau), a(\tau)], \\
	\pi^{\prime}(\mathbf{x}, \tau) &= \mathcal{K}[\phi(\mathbf{x}, \tau), a(\tau), \pi_a(\tau)],
\end{align}
\end{subequations}
where~$\mathcal{D}[\dots]$ is the `drift', and~$\mathcal{K}[\dots]$ the `kernel'. In this case the kernel is~$\mathcal{K}[\dots]=a \nabla^2 \phi - a^3 V_{, \phi}$. \CLa\, solves this first order system of equations using numerical algorithms such as the staggered leapfrog or Verlet integration. We can include the first order metric perturbation terms in the lattice evolution through a modification to the kernel, whereby we add the terms coming from the metric perturbation. These terms are sourced by the fluctuation in the field. The kernel becomes
\begin{align}
	\begin{split}\label{eq:kernelMod}
		& K_{(\delta g)^1}[\dots] = K[\dots] +2\varepsilon(\tau)\mathcal{H}^2(\tau)[\varepsilon(\tau)\\&\qquad\qquad +3-2\eta(\tau)]\delta\phi(\mathbf{x}, \tau)\,.
	\end{split}
\end{align}
For multiple fields, this generalises~\cite{Malik:2008im} to 
\begin{align}
	\begin{split}
		K^i_{(\delta g)^1}[\dots] = K^i[\dots] + \sum_j\left(\frac{a^2\phi_i'\phi_j'}{\mathcal{H}}\right)'\delta\phi_j(\mathbf{x}, \tau)\,,
	\end{split}
\end{align}
where~$K^i$ denotes the kernel for each field~$\phi_i$. The first-order metric perturbation therefore provides an additional coupling between fields, and can be viewed an effective mass matrix (see also~\cite{Caravano:2024xsb,Jamieson:2025ngu}). In~\cref{fig:FreezingMode}, we show for the slow-roll example the evolution of a single mode, which now freezes. In~\cref{fig:PSCLvsModMSUSR}, we display the evolution of the power spectrum for the USR model, both with and without the inclusion of the first-order metric perturbation.

\section{Lattice studies of reheating}
\label{app:reheating}

\begin{figure}
	\includegraphics[width=\linewidth]{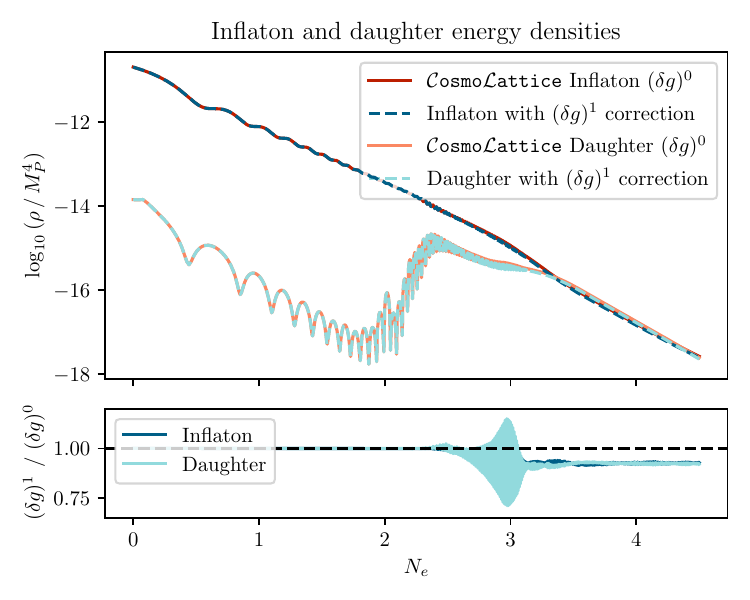}
	\caption{The instantaneous energy density in the inflaton and daughter fields during reheating for \CLa\,~$(\delta g)^0$, and \CLa\, with metric perturbations at first order~$(\delta g)^1$. }\label{fig:Reheating}
\end{figure}

\begin{figure}
	\includegraphics[width=\linewidth]{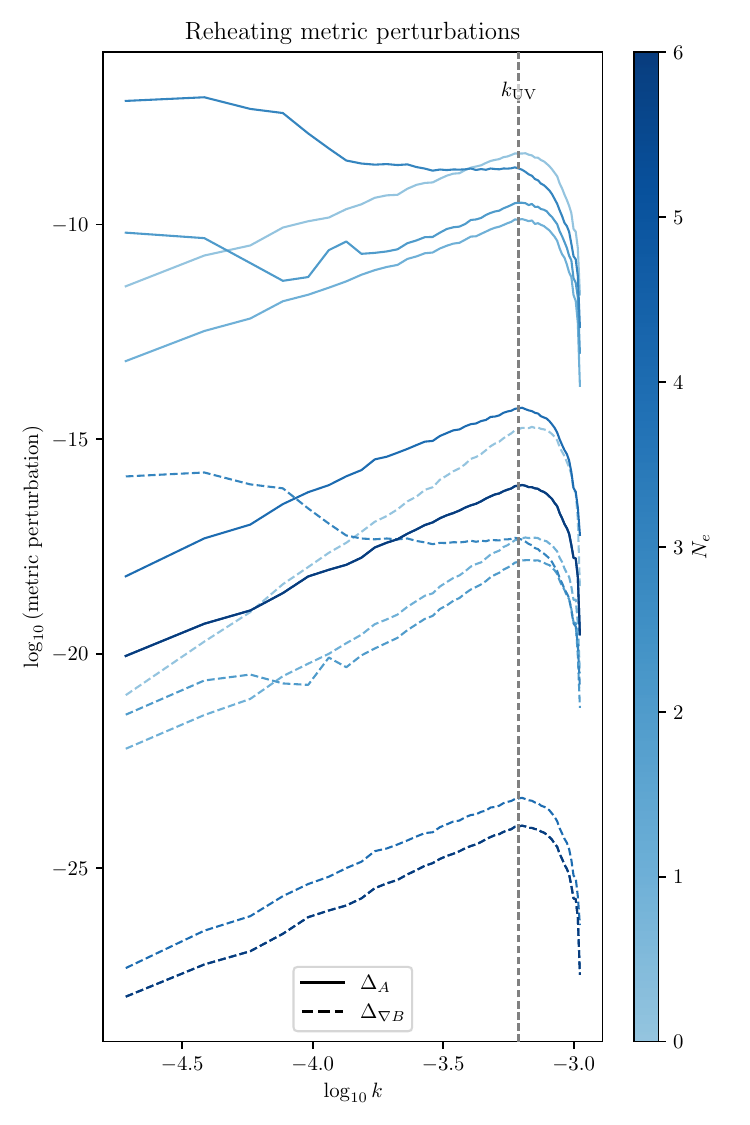}
	\caption{The magnitude of the first order metric perturbations for the reheating model.As in the slow-roll and USR examples, the metric perturbation remains small, but for our reheating scenario their impact is significantly more limited.}\label{fig:ABReheating}
\end{figure}

As an example of reheating, we consider a model from~\cite{Kallosh:2013hoa, Garcia:2020eof,Garcia:2021iag} defined by
\begin{equation}
	\mathcal{L}\supset \lambda \left[ \sqrt{6} \tanh \left( \frac{\phi}{\sqrt{6}} \right) \right]^{2} -\frac{\sigma}{2}\phi^2\chi^2\,,
\end{equation}
where~$\phi$ is the inflaton, and~$\chi$ a daughter field into which it can decay, and we set~$\lambda=2.05\times10^{-11}$, and~$\sigma/\lambda=10^6$. The end of inflation at~$\ddot{a}=0$ determines~$\phi(t_\mathrm{End})\approx 0.84$, which gives the initial conditions for the lattice simulation. In~\cref{fig:Reheating}, we show that including the first order metric perturbation has virtually no effect on the evolution of the instantaneous energy density of the two fields. This happens in spite of the fact that metric perturbations as shown in \cref{fig:ABReheating} are of comparable size as in the~$\lambda\phi^4$ model.

\section{Performant integration of the MS equation} \label{app:ms-numerics}

The MS equation~\cref{generalMS} is numerically demanding for two reasons. It is stiff in the subhorizon regime, where the~$\sim(k/aH)^2$ source dominates and the characteristic eigenvalues of the linearised system span many orders of magnitude across a typical integration window. It is also globally oscillatory at the Bunch--Davies vacuum initial conditions.

\begin{figure*}
	\centering
	\includegraphics[width=\textwidth]{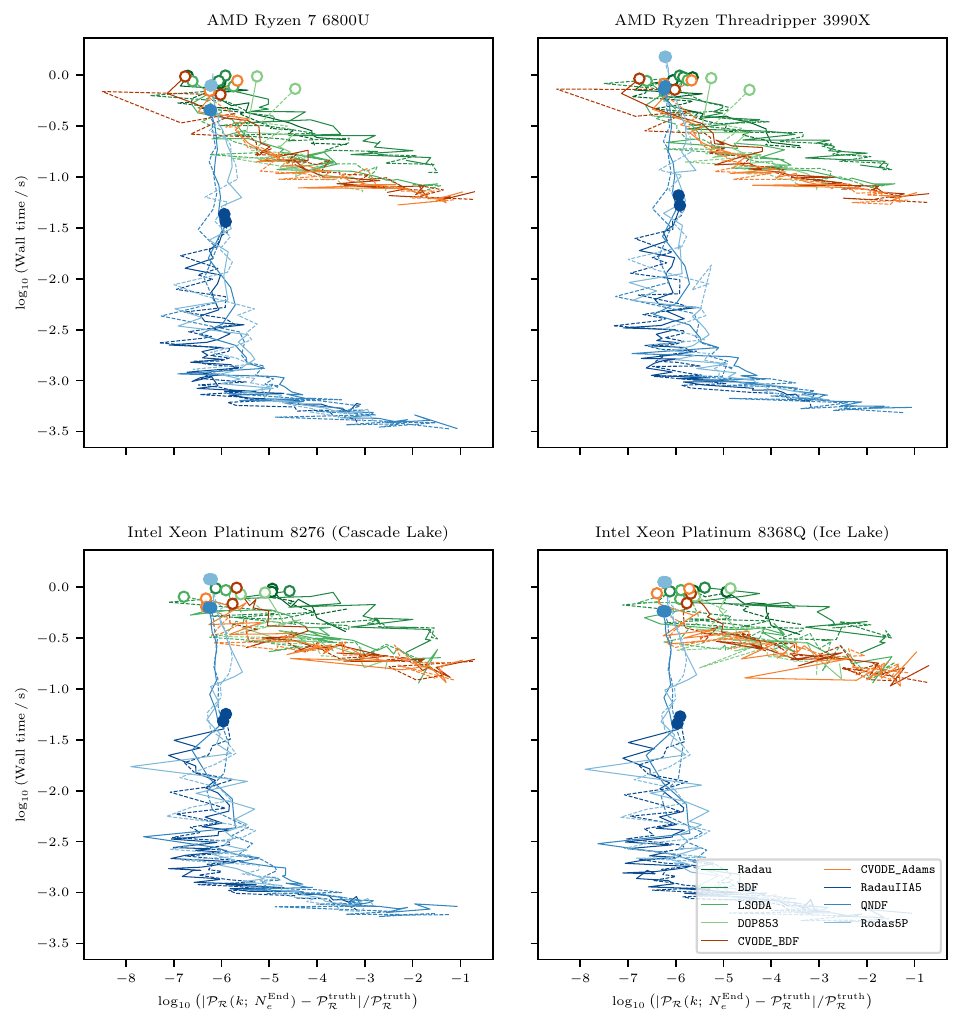}
	\caption{Work-precision diagram for the nine MS solvers at~$k\approx 4.1\times 10^{-9}$, with each solver traced across~$50$ \texttt{rtol} values; solid =~\cref{generalMS}, dashed =~\cref{deformedMS}, and a filled (open) end circle marks where the tightest \texttt{rtol} is (is not) reached. \Python solvers are drawn in green, \Sundials in orange, \Julia in blue. The four panels share common axes and span four CPU architectures, clockwise from top left: a laptop (AMD Ryzen 7 6800U), a workstation (AMD Ryzen Threadripper 3990X), and two server nodes (Intel Xeon Platinum 8368Q, Ice Lake; and 8276, Cascade Lake). The Pareto ordering is preserved across hardware, with absolute wall times shifting only slightly.}
	\label{fig:msPareto}
\end{figure*}

In~\cite{Barker:2024mpz} we used \Scipy's \solver{Radau} implementation as our MS solver. For this work, and to inform future research, we benchmarked the nine most relevant ODE solvers spanning three language ecosystems against both~\cref{generalMS} and its deformed counterpart~\cref{deformedMS}, evaluated on the USR model of~\cref{Jordan} -- the most demanding background appearing anywhere in the present letter. The solver invocations and the language conventions for their five common arguments are given in~\cref{tab:msCalls,tab:msArgs}. The results are shown in~\cref{fig:msPareto}.

Each~$k$ mode is integrated from~$k/aH=20$, \ie approximately~$3$ e-folds inside the horizon, to~$7$ e-folds after horizon crossing (clipped to end of inflation). The Bunch--Davies initial conditions are split into two independent rotations, and the spectrum is reconstructed after solution. The background values of~$\varepsilon$,~$\eta$,~$\dot\phi/\mathcal{H}$ and~$\mathcal{H}$ are integrated once with \Scipy \solver{DOP853} at~$\texttt{rtol} = \texttt{atol} = 10^{-12}$ and passed to every solver as a cubic-spline interpolant. Tolerances are swept over~$50$ values of~\texttt{rtol} logarithmically spaced from~$10^{-3}$ to~$10^{-12}$, with~$\texttt{atol} = 10^{-3}\,\texttt{rtol}$. The ground-truth value~$\mathcal{P}^{\mathrm{truth}}_{\mathcal{R}}$ used as the reference on the horizontal axis of~\cref{fig:msPareto} is computed separately for each of~\cref{generalMS} and~\cref{deformedMS} as the unweighted mean of the three \Julia methods, since these reliably reach the tightest tolerance~$\texttt{rtol}=10^{-12}$. The \Python and \Sundials solvers run in a process pool with a~$1\,\mathrm{s}$ budget per solution: once a solve at some \texttt{rtol} exceeds the budget, all tighter \texttt{rtol}s for that method are assumed to similarly fail. The \Julia solvers run in a single long-lived subprocess that covers all \texttt{rtol}s and equations internally, paying their just-in-time compilation cost only once.

As shown in~\cref{fig:msPareto}, the three \Julia methods adapted for stiff systems (\solver{RadauIIA5}, \solver{QNDF}, \solver{Rodas5P}) are the most performant on all four CPU architectures tested, achieving any chosen accuracy at one to two orders of magnitude lower wall time than the \Python/\Sundials methods, with \solver{RadauIIA5} holding a slight edge over the other two. Among the \Python solvers, \Scipy's \solver{Radau} remains the most reliable choice at the tightest tolerances (hence its previous use in~\cite{Barker:2024mpz}). Though not shown, this ranking is found to be mostly consistent across~$k$ values for the current class of problems. We do not make a comprehensive survey of the other inflationary parameters or consider alternative inflationary backgrounds. For each~$k$ mode, the USR spectra plotted in this letter (e.g.~the right column of~\cref{fig:observablesUSR}) are taken as the unweighted mean of the three \Julia methods, each integrated at~$\texttt{rtol}=10^{-12}$; the slow-roll spectra are instead integrated with \Scipy's \solver{Radau}.

In summary, we are able to provide the following heuristic recommendation for future work:
\begin{center}
	\emph{For performant, science-grade integration of the Mukhanov--Sasaki equation at the time of writing, one should prefer \Julia's \solver{RadauIIA5} method, or if it is necessary to use \Python, \Scipy's \solver{Radau} method.}
\end{center}
The full benchmark data and code are available at~\cite{supp_materials}.

\begin{table*}
\centering
\caption{Solver invocations for the nine MS integrators benchmarked in~\cref{fig:msPareto}.}
\label{tab:msCalls}
\begin{tabular}{lll}
\hline
Ecosystem & Method & Call \\
\hline
\Python   & \solver{Radau}        & \lstinline|solve_ivp(f, tspan, y0, method='Radau',   rtol=rtol, atol=atol)| \\
\Python   & \solver{BDF}          & \lstinline|solve_ivp(f, tspan, y0, method='BDF',     rtol=rtol, atol=atol)| \\
\Python   & \solver{LSODA}        & \lstinline|solve_ivp(f, tspan, y0, method='LSODA',   rtol=rtol, atol=atol)| \\
\Python   & \solver{DOP853}       & \lstinline|solve_ivp(f, tspan, y0, method='DOP853',  rtol=rtol, atol=atol)| \\
\Sundials & \solver{CVODE\_BDF}   & \lstinline|CVODE(f, method='BDF',   rtol=rtol, atol=atol).solve(tspan, y0)| \\
\Sundials & \solver{CVODE\_Adams} & \lstinline|CVODE(f, method='Adams', rtol=rtol, atol=atol).solve(tspan, y0)| \\
\Julia    & \solver{RadauIIA5}    & \lstinline|solve(ODEProblem(f, y0, tspan), RadauIIA5(); reltol=rtol, abstol=atol)| \\
\Julia    & \solver{QNDF}         & \lstinline|solve(ODEProblem(f, y0, tspan), QNDF();      reltol=rtol, abstol=atol)| \\
\Julia    & \solver{Rodas5P}      & \lstinline|solve(ODEProblem(f, y0, tspan), Rodas5P();   reltol=rtol, abstol=atol)| \\
\hline
\end{tabular}
\end{table*}

\begin{table*}
\centering
\caption{\Python and \Julia conventions for the five arguments appearing in~\cref{tab:msCalls}.}
\label{tab:msArgs}
\begin{tabular}{lll}
\hline
Name & \Python & \Julia \\
\hline
\texttt{f}     & \texttt{f(t, y)} returns~$[\dot y_0,\dot y_1]$ (\Scipy); \texttt{f(t, y, yp)} writes \texttt{yp} (\Sksundae) & \texttt{f!(du, u, p, t)} writes \texttt{du} \\
\texttt{tspan} & list \texttt{[Ni, Nf]}     & tuple \texttt{(Ni, Nf)} \\
\texttt{y0}    & \texttt{np.array([f0, fp0])} & \texttt{[f0, fp0]} (\texttt{Vector\{Float64\}}) \\
\texttt{rtol}  & \texttt{float}, kwarg \texttt{rtol} & \texttt{Float64}, kwarg \texttt{reltol} \\
\texttt{atol}  & \texttt{float}, kwarg \texttt{atol} & \texttt{Float64}, kwarg \texttt{abstol} \\
\hline
\end{tabular}
\end{table*}

\bibliography{RefsInspire,RefsManual}

\end{document}